%% file: main.tex
\documentclass[a4paper,11pt]{article}
\pdfoutput=1 

\usepackage{jinstpub}
\usepackage{orcidlink}
\usepackage{lineno}
\usepackage{siunitx}

\title{First Demonstration of a Combined Light and Charge Pixel Readout on the Anode Plane of a LArTPC}

\input{authorlist.tex}

\collaboration{\includegraphics[height=17mm]{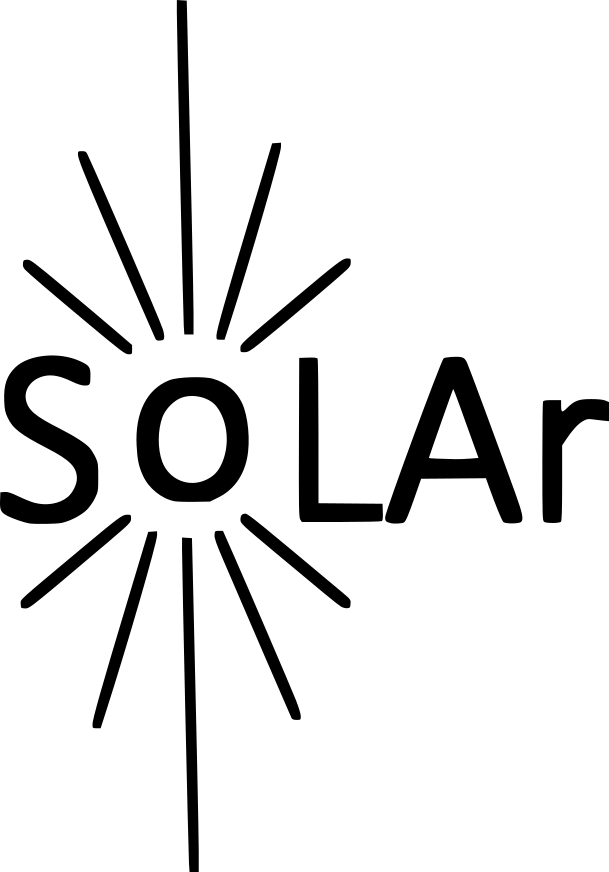}\\[6pt]
The SoLAr Collaboration}

\emailAdd{saba.parsa@cern.ch}

\input{abstract.tex}

\keywords{Noble liquid detectors (scintillation, ionization, single-phase), time projection chambers, neutrino detectors}

\collaboration{SoLAr Collaboration}

\begin{document}
\maketitle

\flushbottom

\input{section1_introduction.tex}
\input{section3_design.tex}
\input{section4_cryo.tex}
\input{section5_cosmic_run.tex}
\input{section6_conclusion.tex}

\section*{Acknowledgements}

This document was prepared by the SoLAr collaboration.
The project has received funding from the European Union’s Horizon 2020 Research and Innovation programme under GA no 101004761, 
by the Italian Ministry for Research and University (MUR) under Grant Progetto Dipartimenti di Eccellenza 2023-2027, and by
the Swiss National Science Foundation. It has also been supported by several of the collaborating institutions:
CIEMAT;
Lawrence Berkeley National Laboratory;
University of Bern;
University of Manchester;
University of Milano-Bicocca and INFN Sezione di Milano-Bicocca.

\bibliographystyle{JHEP}
\bibliography{bib}

\end{document}

%% file: authorlist.tex
\author[13]{N.~Anfimov,}
\author[6,10]{A. Branca,}
\author[1]{J.~B\"{u}rgi,} 
\author[1]{L.~Calivers,} 

\author[3]{C.~Cuesta,} 

\author[1]{R.~Diurba,} 
\author[5]{P.~Dunne,} 
\author[7]{D.~A.~Dwyer,} 

\author[9]{J.~J.~Evans,} 
\author[12]{A.~C.~Ezeribe,} 

\author[1]{A.~Gauch,} 
\author[3]{I.~Gil-Botella,} 
\author[2,7]{S.~Greenberg,} 
\author[6,10]{D.~Guffanti,}  

\author[7]{A.~Karcher,} 
\author[1]{I.~Kreslo,} 
\author[1]{J.~Kunzmann,} 

\author[9]{N.~Lane,} 

\author[3]{S.~Manthey~Corchado,} 
\author[11]{N.~McConkey,} 

\author[5,9]{A.~Navrer-Agasson,}  

\author[1]{S.~Parsa,} 

\author[9]{G.~Ruiz~Ferreira,} 
\author[8]{B.~Russell,} 

\author[13]{A.~Selyunin,}
\author[5,9]{S.~S\"{o}ldner-Rembold,} 
\author[4]{A.~M.~Szelc,} 

\author[5]{A.~Tapper,} 
\author[6,10]{F.~Terranova,} 
\author[1]{C.~Tognina,} 

\author[4]{G.~V.~Stenico,} 

\author[1]{M.~Weber,} 

\author[5]{and I.~Xiotidis} 

\affiliation[1]{University of Bern, CH-3012 Bern, Switzerland}
\affiliation[2]{University of California Berkeley, Berkeley, CA 94720, USA}
\affiliation[3]{CIEMAT, Centro de Investigaciones Energ{\'e}ticas, Medioambientales y Tecnol{\'o}gicas, E-28040 Madrid, Spain}
\affiliation[4]{University of Edinburgh, Edinburgh EH9 3FD, United Kingdom}

\affiliation[5]{Imperial College of Science, Technology and Medicine, London SW7 2BZ, United Kingdom}
\affiliation[6]{Istituto Nazionale di Fisica Nucleare Sezione di Milano Bicocca, 3 - I-20126 Milano, Italy}

\affiliation[7]{Lawrence Berkeley National Laboratory, Berkeley, CA 94720, USA}

\affiliation[8]{Massachusetts Institute of Technology, Massachusetts, MA 02139, USA}
\affiliation[9]{University of Manchester, Manchester M13 9PL, United Kingdom}
\affiliation[10]{Universit{\`a} di Milano-Bicocca, I-20126 Milano, Italy}

\affiliation[11]{Queen Mary University of London, London E1 4NS, United Kingdom}
\affiliation[12]{University of Sheffield, Sheffield S3 7RH, United Kingdom}
\affiliation[13]{Visitor}

%% file: abstract.tex
\abstract{The novel SoLAr concept aims to extend sensitivities of liquid-argon neutrino detectors down to the MeV scale for next-generation detectors. SoLAr plans to accomplish this with a liquid-argon time projection chamber that employs an anode plane with dual charge and light readout, which enables precision matching of light and charge signals for data acquisition and reconstruction purposes. We present the results of a first demonstration of the SoLAr detector concept with a small-scale prototype detector integrating a pixel-based charge readout and silicon photomultipliers on a shared printed circuit board. We discuss the design of the prototype, and its operation and performance, highlighting the capability of such a detector design.}

%% file: section1_introduction.tex
\section{Introduction}
\label{sec_intro}
The next generation of deep underground detectors, namely DUNE~\cite{DUNETDRVol1}, will enable a substantial leap in the sensitivity of neutrino physics experiment through their detectors' mass and precision. Their baseline design is optimized for the detection of GeV-scale neutrinos produced by a particle beam with the goal to measure oscillation parameters with unprecedented precision~\cite{DUNE:2020jqi}. 
In addition, the large mass and the underground location of the detectors provide a unique opportunity to expand their sensitivity to lower energy neutrinos, with energies at the MeV-scale, originating from natural nuclear processes.

Solar neutrinos provide an avenue for these future detectors to explore, primarily to detect neutrinos generated from the \textit{hep} branch~\cite{2004HEPReview, SNO2020}. Its discovery directly impacts the modeling of our Sun and the modeling of stellar evolution throughout the Universe. The application of liquid-argon detectors to solar-neutrino physics has been explored previously~\cite{Capozzi, IGil-Botella_2004, Fogli:2006fu}. Such a detector was found to need an energy resolution at the MeV-scale of $7 \%$ to measure solar neutrinos~\cite{Capozzi}, to have excellent background rejection capabilities, and to have reasonably low data rates. These requirements are essential to successfully discriminate the higher-energy tail of the \textit{hep} neutrino spectrum from the dominant $^8$B neutrino spectrum. Supernova explosions eject most of their energy in form of neutrinos with energies similar to those of \textit{hep} neutrinos. A detector that is able to observe \textit{hep} neutrinos is by construction also the most precise experiment to detect supernova explosions. Combining the data with visible, X-ray, $\gamma$-ray, and gravitational-wave information will provide a complete picture of the supernova 
collapse~\cite{DUNE:2020zfm}.

One significant challenge is related to readout and computing. The whole DUNE Far Detector site needs to collect petabytes of light and charge data per year to accomplish its calibration and neutrino programs from natural sources~\cite{DUNETDRVol1}. Providing a solution that enables large mass detectors while limiting data rates would significantly increase the feasibility of low-energy searches in liquid-argon detectors. 

The SoLAr (solar neutrinos in liquid argon) collaboration proposes to build multipurpose liquid-argon time projection chambers (LArTPCs) with several kilotons of active mass to detect beam neutrinos, supernova neutrinos and solar neutrinos, in particular those produced in the \textit{hep} process in the Sun~\cite{Parsa:2022mnj}. The main feature of the SoLAr proposal consists of a pixel-based dual-readout anode plane combining charge and light readout to increase energy resolution, improve background rejection, and decrease data storage requirements. 

The current DUNE Far Detector modules have separate charge readout sensors and large area light detectors that can sit behind the anode or on the cathode~\cite{DUNETDRVol4,DUNE:2023nqi}. The DUNE Near Detector liquid-argon modules have a pixel-based charge collecting anode plane with large area dielectric light collecting modules adjacent to the anode plane within the electric field of the LArTPC~\cite{duneNDCDR}. The SoLAr design uniquely has discrete, pixelated light and charge readout on a single, shared printed circuit board (PCB). The SoLAr detector concept is based on the same detector technology of pixel-based LArTPCs being constructed for the Near Detector site at DUNE~\cite{duneNDCDR}. The pixel-based charge readout is able to do native three-dimensional reconstruction. For the SoLAr detector concept, individual silicon photomultipliers (SiPMs) will be distributed in a uniform pattern among the charge pixels on the anode plane, providing a localized light reconstruction to accompany the signals from the neighboring charge pixels. The development of SiPMs sensitive to vacuum ultraviolet (VUV) light is instrumental to the SoLAr proposal. The combination with charge readout allows for greater sensitivities in reconstruction through light and charge signal matching. It opens possibilities in online triggering by sectorizing the detector and triggering specific pixels only when its neighboring light detector sees a signal.

We aim to use the design concept to build an $\approx 10$~t 
liquid-argon SoLAr-type detector with an active volume of about $1.6\times 2 \times 2.6$~m$^3$. Early planning has found the Boulby Underground Laboratory in the United Kingdom as a suitable location. The Boulby Laboratory is located in a working polyhalite and salt mine in the North East of England
at a depth of $1.1$~km.
The SoLAr detector concept could also be adapted for the instrumentation of future Far Detector modules for DUNE~\cite{Parsa:2022mnj} as part of the DUNE Phase-II programme~\cite{DUNE:2022aul,dune2}. 

In this paper, we describe the design of the first SoLAr prototype TPC and present results from the operation of this prototype at the University of Bern in October 2022.

%% file: section3_design.tex
\section{Experimental Setup}
\subsection{Design of the SoLAr prototype Time Projection Chamber}
\label{sec_design}
The first SoLAr prototype TPC (Prototype-v1) uses a PCB with a pixelated charge readout system using the LArPix chip~\cite{Dwyer:2018phu} and neighboring Hamamatsu VUV SiPMs. The PCB acts as the anode plane and is placed within an electric field to measure the ionization charge in the liquid argon. The liquid argon VUV scintillation light from cosmic-ray muons is measured by the SiPMs. 
The dimensions of the TPC are
$11.8 \times 10.8$~cm$^2$ in the $x$ and $y$ direction, with a sensitive anode area of $7 \times 7$~cm$^2$, and $5$~cm in the drift direction ($z$). The choice is driven by the cryostat inner volume geometry, which has a diameter of $14$~cm. This size is sufficient to demonstrate the SoLAr concept by collecting light and charge from cosmic rays on the same plane. 

\begin{figure}[htb]
\centering 
\includegraphics[width=.47\textwidth]{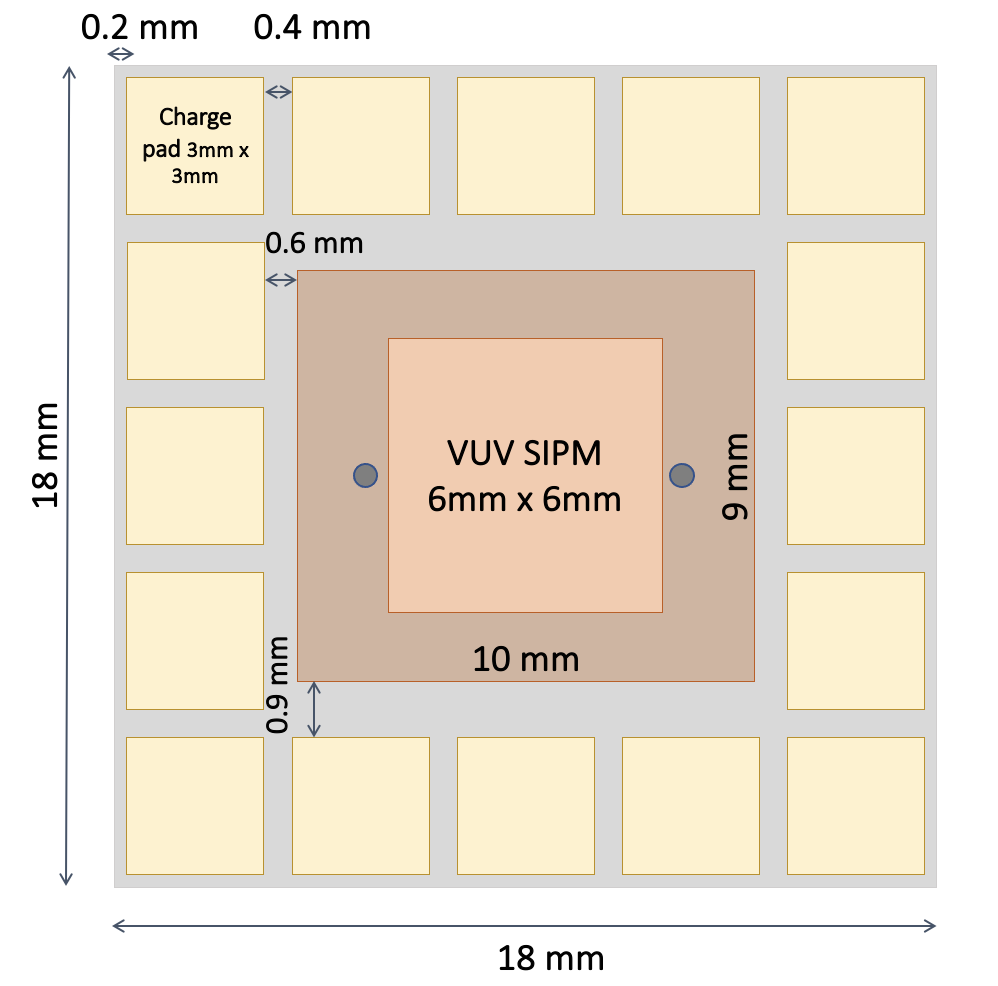}
\qquad
\includegraphics[width=.47\textwidth]{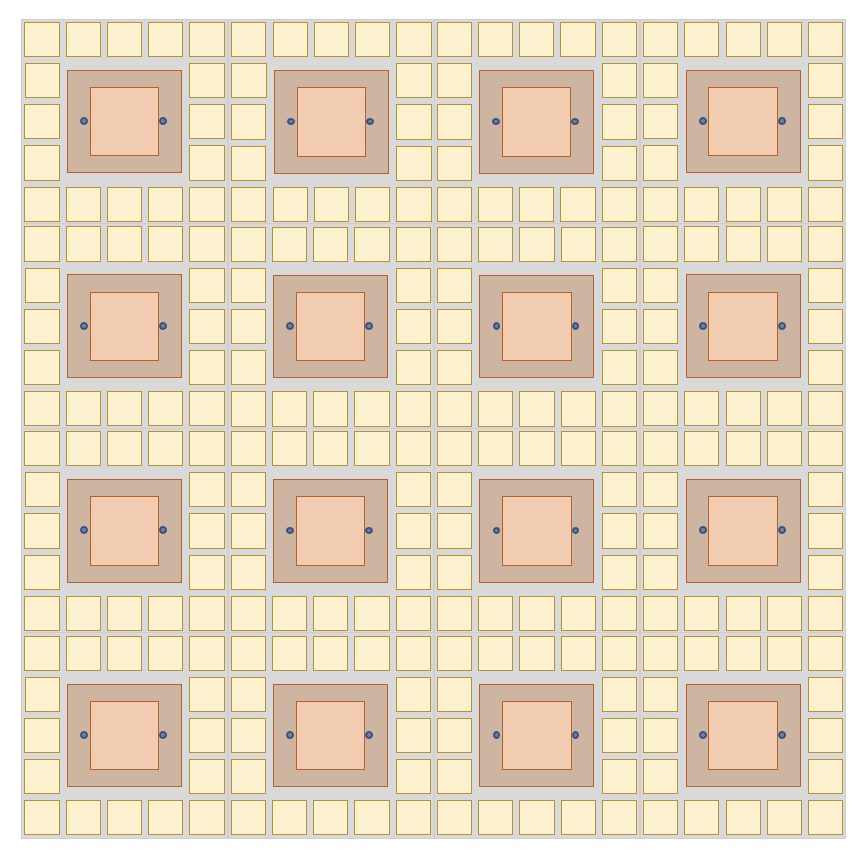}
\caption{(a) Schematic of a single cell within the sensitive area of the LArTPC. Charge-collecting pixels are located directly next to the pin of each SiPM; (b) the complete v1 prototype with all 16 cells.}\label{fig:cells}
          \vspace{-90mm}
\makebox[\textwidth][l]{\hspace{1.1cm}(a) \hspace{7.4cm} (b)}\\ \vspace{90mm}
\end{figure}

The active area of the LArTPC is divided into a grid of $16$ identical cells. Each cell consists of a VUV SiPM (Hamamatsu S13370-6050CN, QE($\lambda=128 nm$) $\approx$ 15\%) and 16 charge-collection pads,
resulting in a total of $256$ pixels~(see Figure~\ref{fig:cells}). 
The cathode is a $1$~cm thick metal plate with smooth golden surface and rounded corners to prevent discharges to the cryostat walls for potentials up to $10$~kV. The side panel PCBs on the four walls function as electric field shaping surfaces with nine metalized bands linked through a resistor chain to create a homogeneous electric field along the electrons' drift line from the cathode to the anode.  

\begin{figure}[!hb]
\centering 
\includegraphics[width=0.7\textwidth, angle=-90]{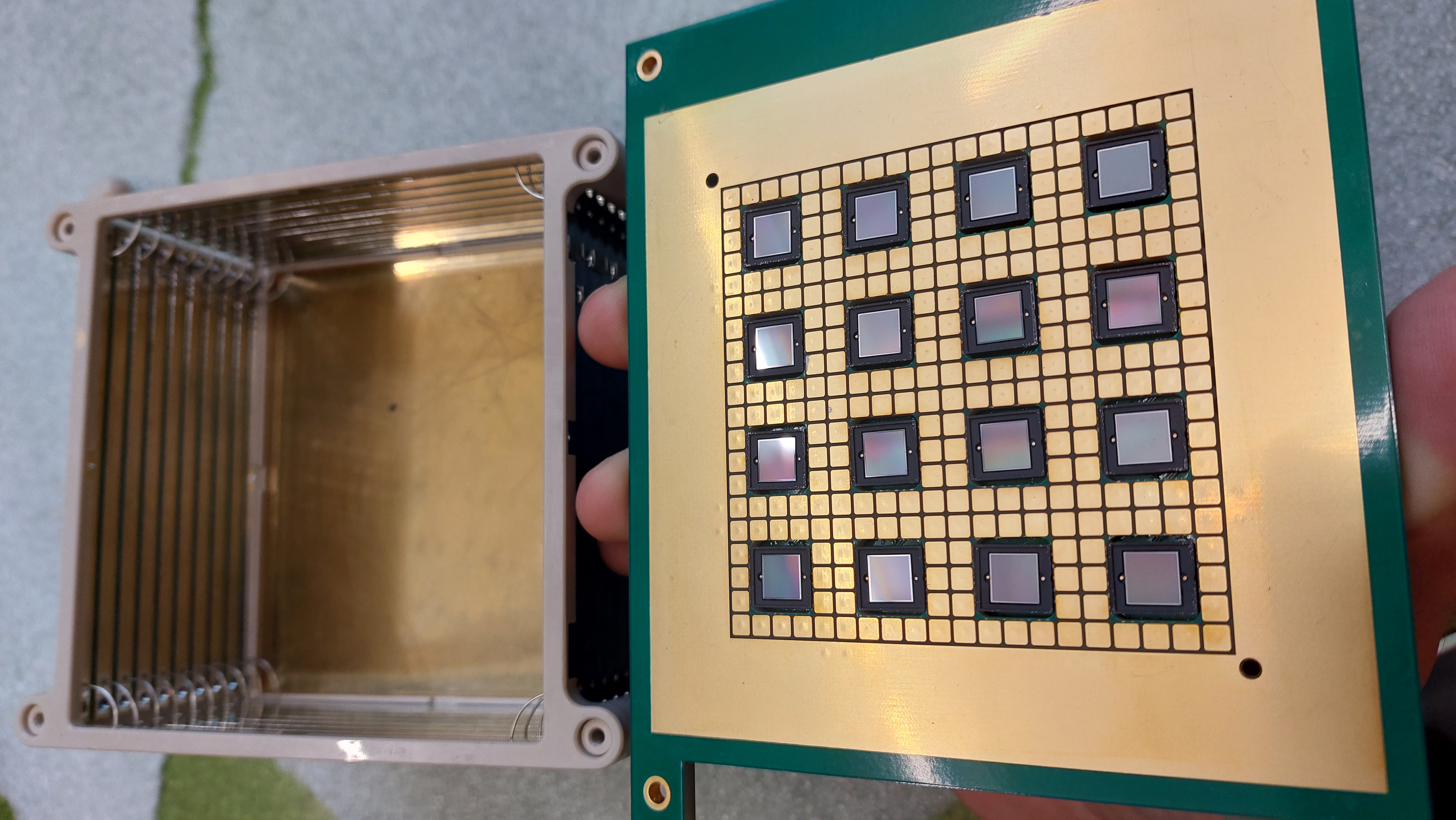} 
\hskip 1.5cm
\includegraphics[width=0.7\textwidth, angle=-90]{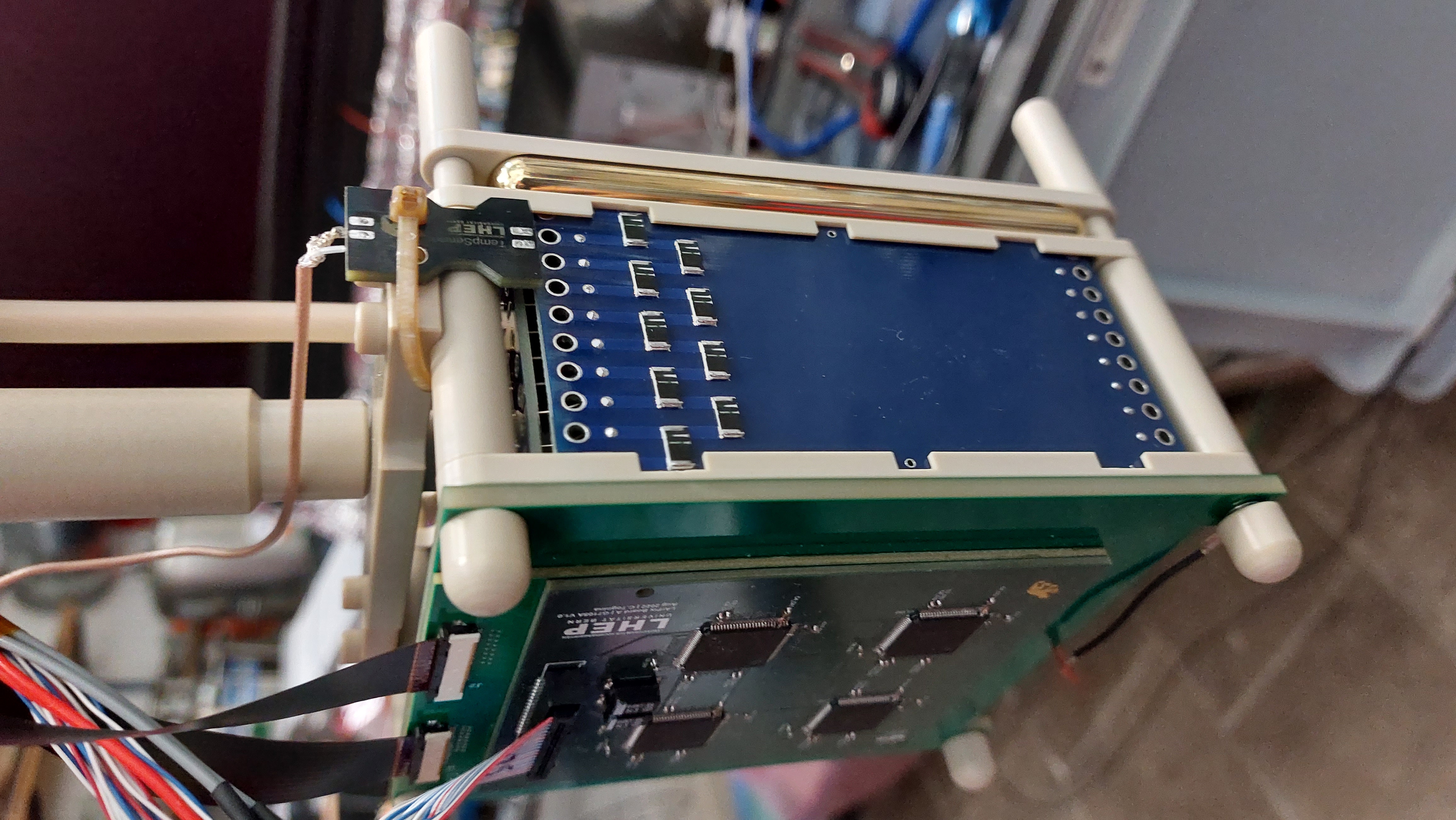}
\qquad

\caption{(left) Interior of the SoLAr-v1 TPC time projection chamber with the cathode plate, the side panels for $E$-field shaping and the readout anode plate containing the SiPMs and charge pixels. 
(right) Fully assembled SoLAr-v1 TPC attached to its support structure. The back side of the anode plate with the four LArPix chips and the back side of one of the E-field shaping panels with a resistor chain is also visible.}
\label{fig:tpc1}

\end{figure}

Figure~\ref{fig:tpc1} shows the interior of the TPC with the cathode plate, side panel PCBs, and the readout anode plate. 
The assembled TPC connected to a hanging fixture is shown
in Figure~\ref{fig:tpc1}. The rounded corner rods extending out of the TPC volume are mechanical guides for insertion into the cryostat and for adjusting the position and distances of the TPC from the cryostat's inner walls.

The VUV SiPMs collect the photons in Geiger mode and transfer the analog light waveform signal to a cold pre-amplifier stage over a short ($20$~cm) double-layered flex PCB. The  signal is then transferred via a feed-through PCB using SAMTEC micro-coaxial cables to a variable gain amplification (VGA) unit. The differential amplified signal at this stage is transferred via a twisted-pair ribbon cable to an analog-to-digital converter (ADC) unit.

The charge collected on the pixel pads is read out with the help of a LArPix chip, a cryo-rated ASIC for pixelated liquid-argon TPCs, developed by LBNL~\cite{Dwyer:2018phu}. Four LArPix chips are used to read out the $256$ pixels on the anode plane. The LArPix chip provides a self-triggering digitization and multiplexed data transfer to a readout hub in warm, referred to as the PACMAN. 
\begin{figure}[htb!]
    \centering
    \includegraphics[width=0.99\textwidth]{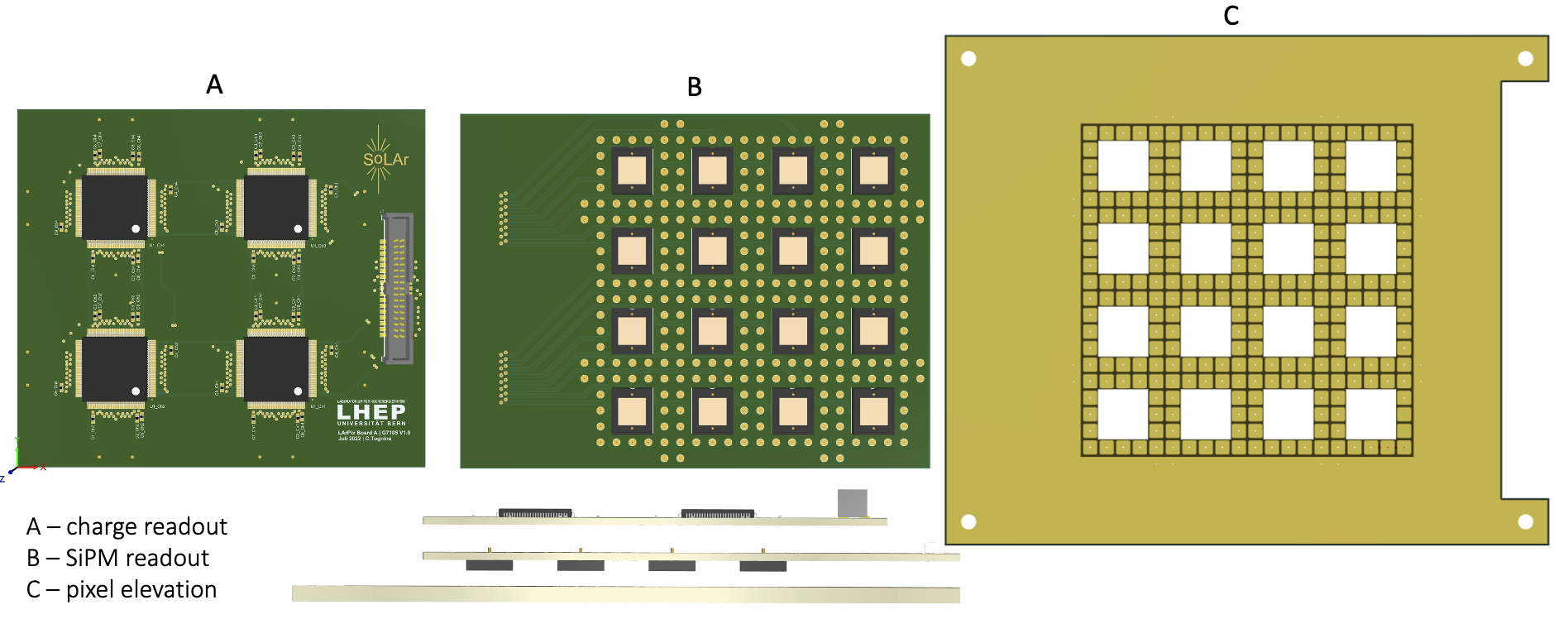}
    \caption{Top and side-view of the anode readout plane three-layer stack: charge (A) and 
    SiPM readout (B), and pixel elevation (C).}
    \label{fig:pcb_stackup}
\end{figure}
The anode readout plane consists of three layers of PCBs mounted on top of each other, as shown in Figure~\ref{fig:pcb_stackup}. The charge collection pads are located on the innermost layer, which contains cutout holes to accommodate the SiPMs soldered on the second layer. This configuration ensures that the surfaces of the ceramic-packaged SiPMs are level with the charge collection pixels. This alignment minimizes electric field deformations near the edges of the SiPMs and charge collection pads.
The second layer hosts the SiPMs and their readout traces to the connectors. It also contains a network of traces that transmits the charge signals to the outermost layer, which hosts the LArPix readout chips and charge readout connector.

The SiPMs operate at a bias voltage of $56$~V at room temperature and $46$~V at liquid-argon temperature with an over voltage between $3$~V and $4$~V. The setup is designed to provide ground offset bias voltages on the SiPMs, such that the top surface of the SiPMs, facing the TPC interior, could be set to a negative ground offset voltage compared to the common ground of the charge pads. In this way, the potential on the SiPMs surface can shape the electric fields toward the pixels, potentially increasing the charge collection efficiency.  

\subsection{Electric field simulation}
\begin{figure}[htbp]
    \centering
    \includegraphics[width=0.49\textwidth]{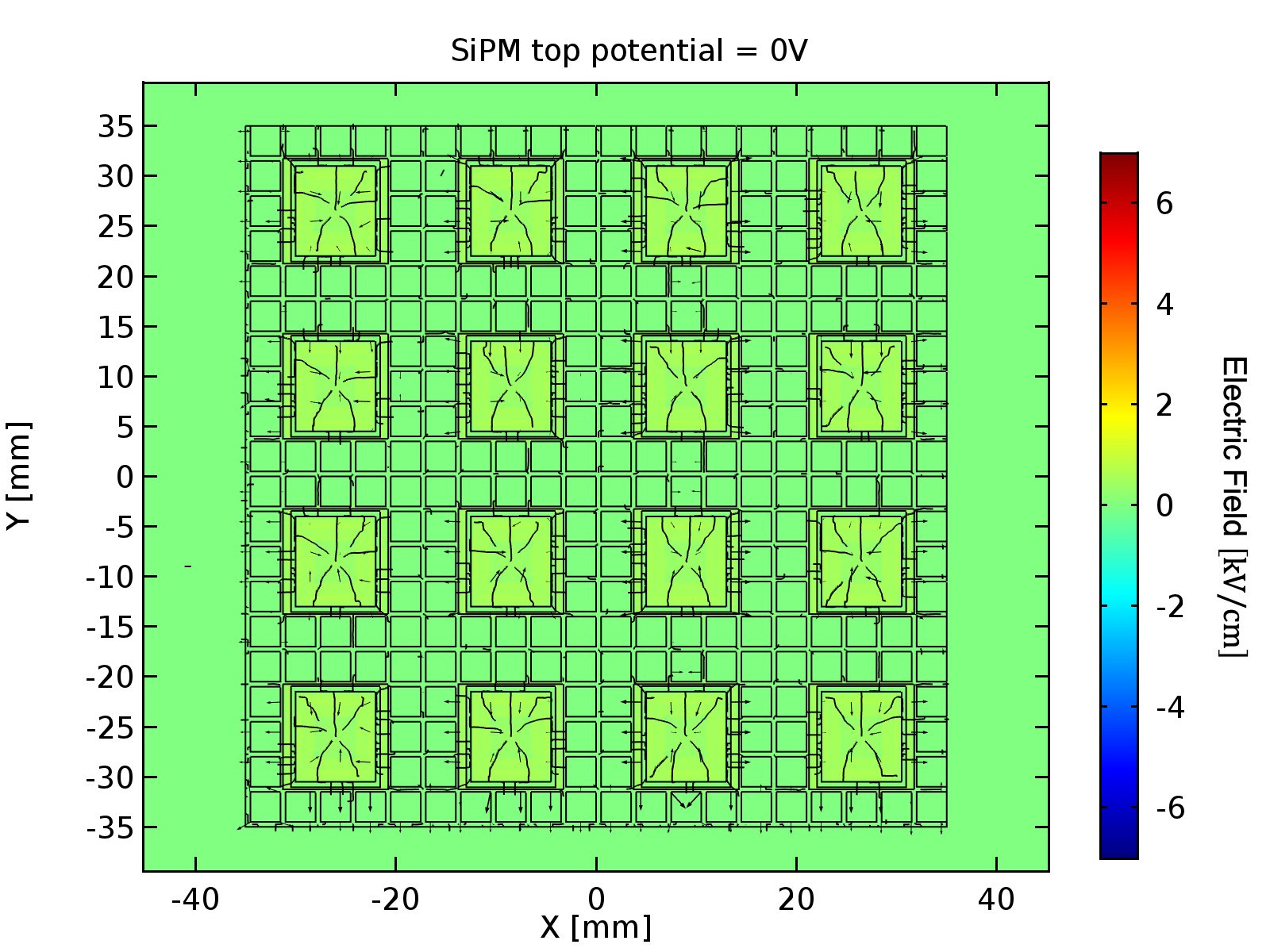}
    \includegraphics[width=0.49\textwidth]{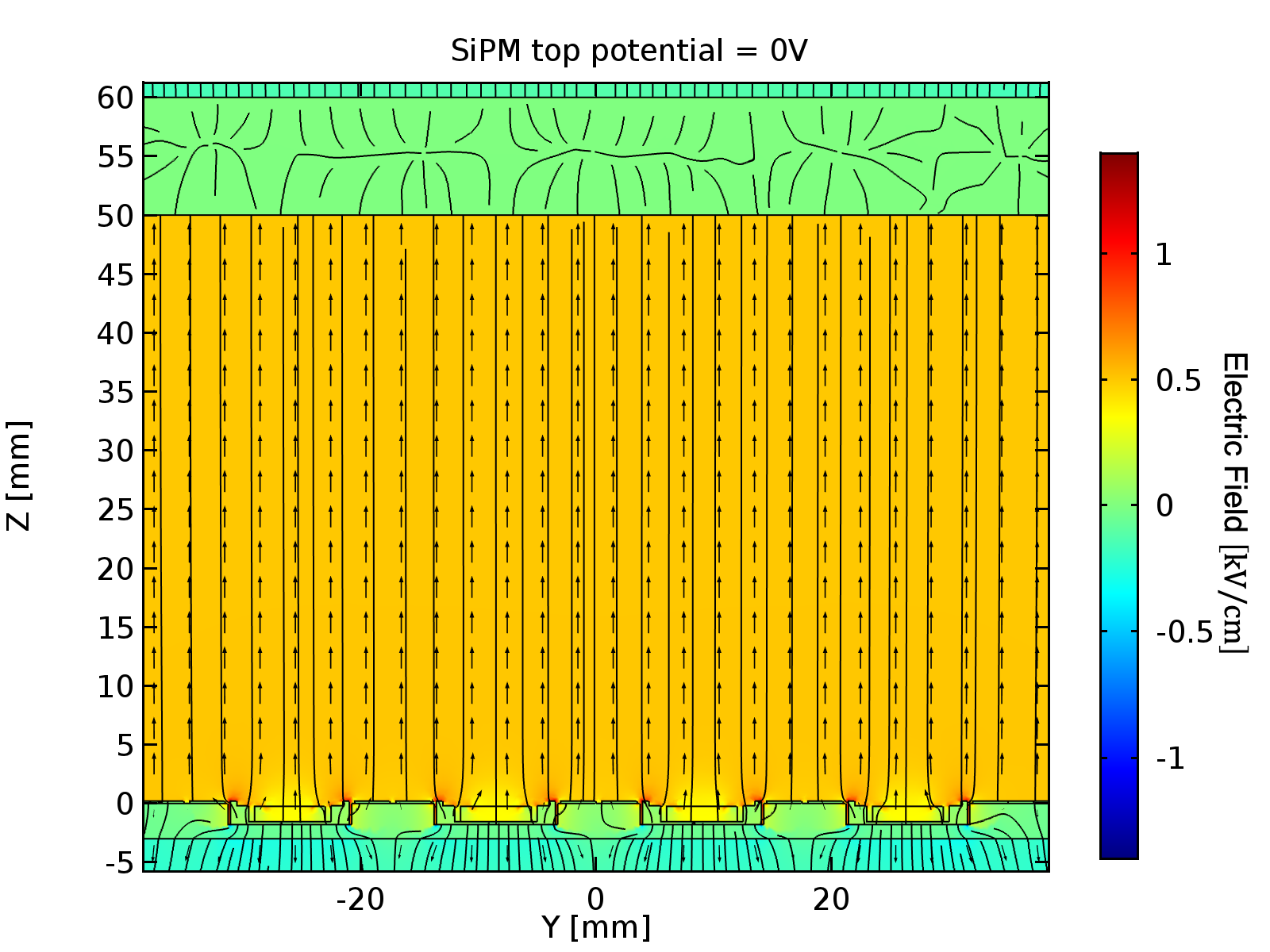}
    \includegraphics[width=0.49\textwidth]{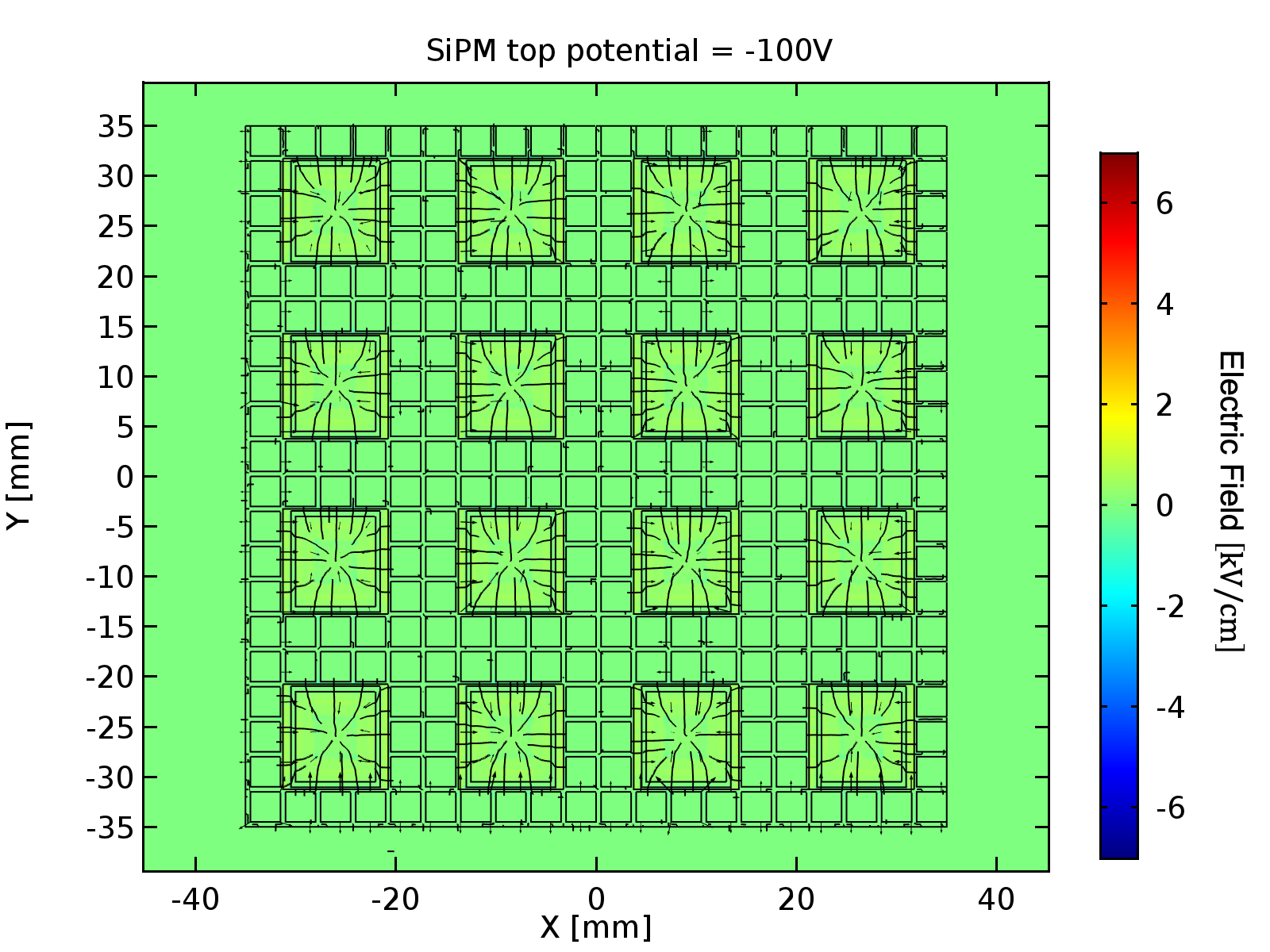}
    \includegraphics[width=0.49\textwidth]{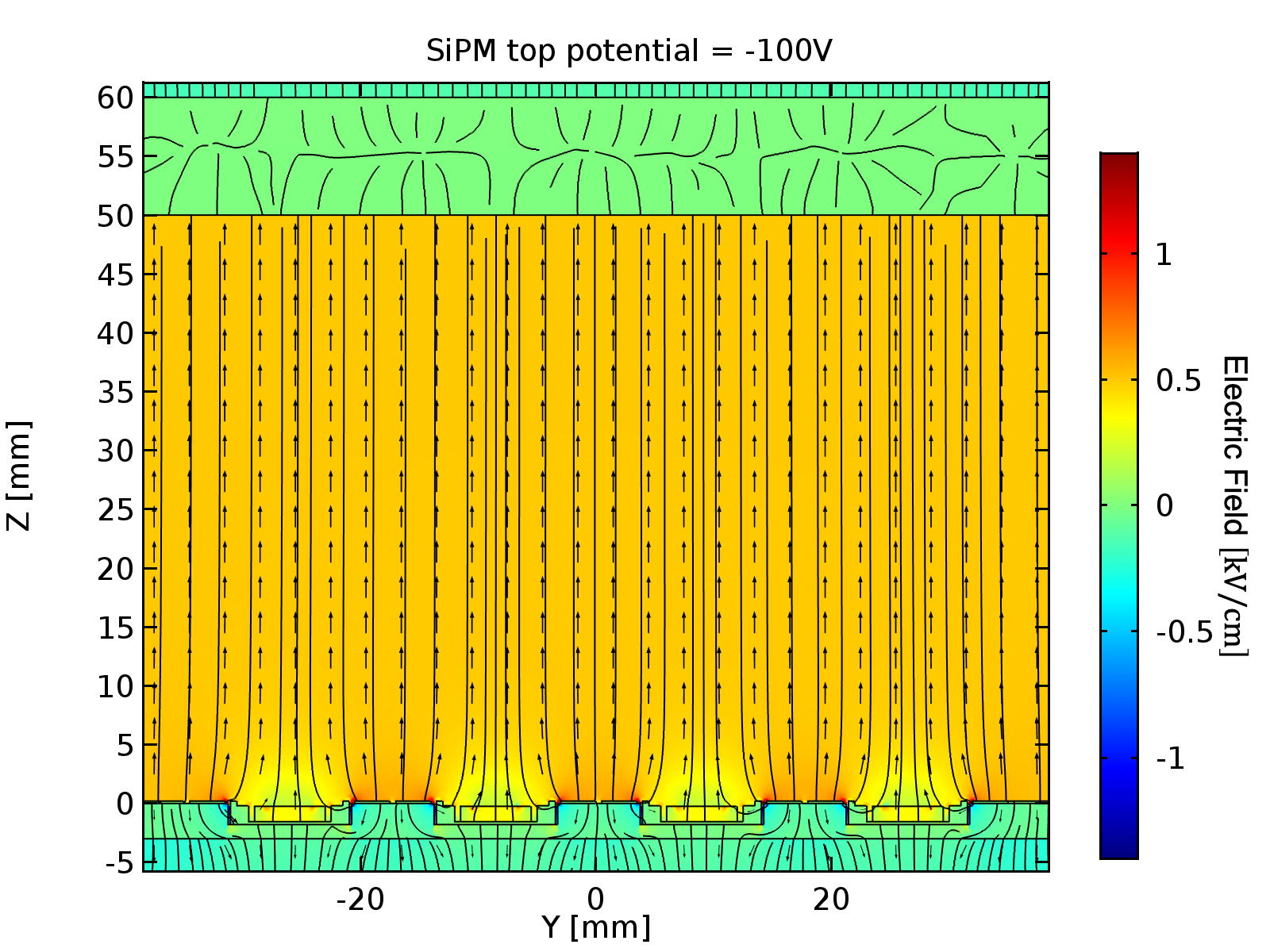}
    \caption{
    COMSOL numerical evaluation showing the direction and magnitude of the electric field on the SoLAr TPC. (Top) SiPM top surface potential is set to ground.
    (Bottom) SiPM top surface potential is set to $-100$~V. (Left) Surface directly on top of the anode plane. (Right) Cross-section of the TPC cutting through the middle of the SiPMs and pixels. }
    \label{fig:comsol}
\end{figure}

To understand the impact of design choices on the charge collection, the electric field in proximity of the anode was evaluated numerically using the COMSOL software~\cite{comsol} for different ground offset values. Raising the ground offset has to be done carefully since a strong electric field inside the SiPMs may effect their operation and cause damages. Moreover, if the ground offset is too large, field lines start to come out of the SiPMs and into the pixels. Figure~\ref{fig:comsol} shows the results for the electric field lines in two different configurations: ground offset set to 0~V, representing the nominal conditions of the run, and -100~V, representing the maximum ground offset of data taking with the SiPM on the three-layered PCB. Data was taken at 0~V (nominal), -25~V, -50~V, -75~V, and -100~V with light data results discussed in Section~\ref{sec_run}.

%% file: section4_cryo.tex
\subsection{Cryogenics system}
\label{sec_cryo}
\begin{figure}[htbp]
\centering 
\includegraphics[height = 10cm]{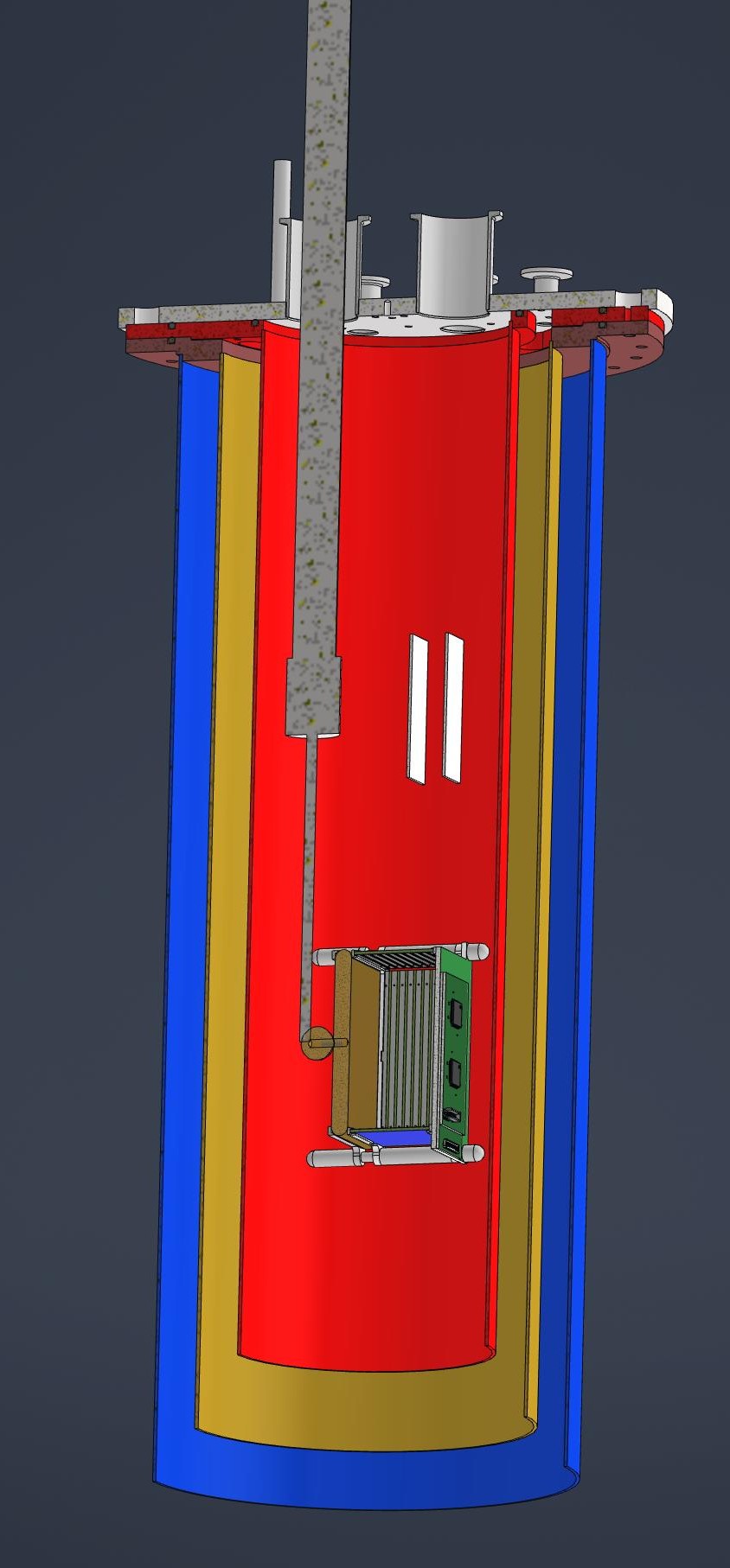}
\hskip 1.5cm
\includegraphics[height =10cm]{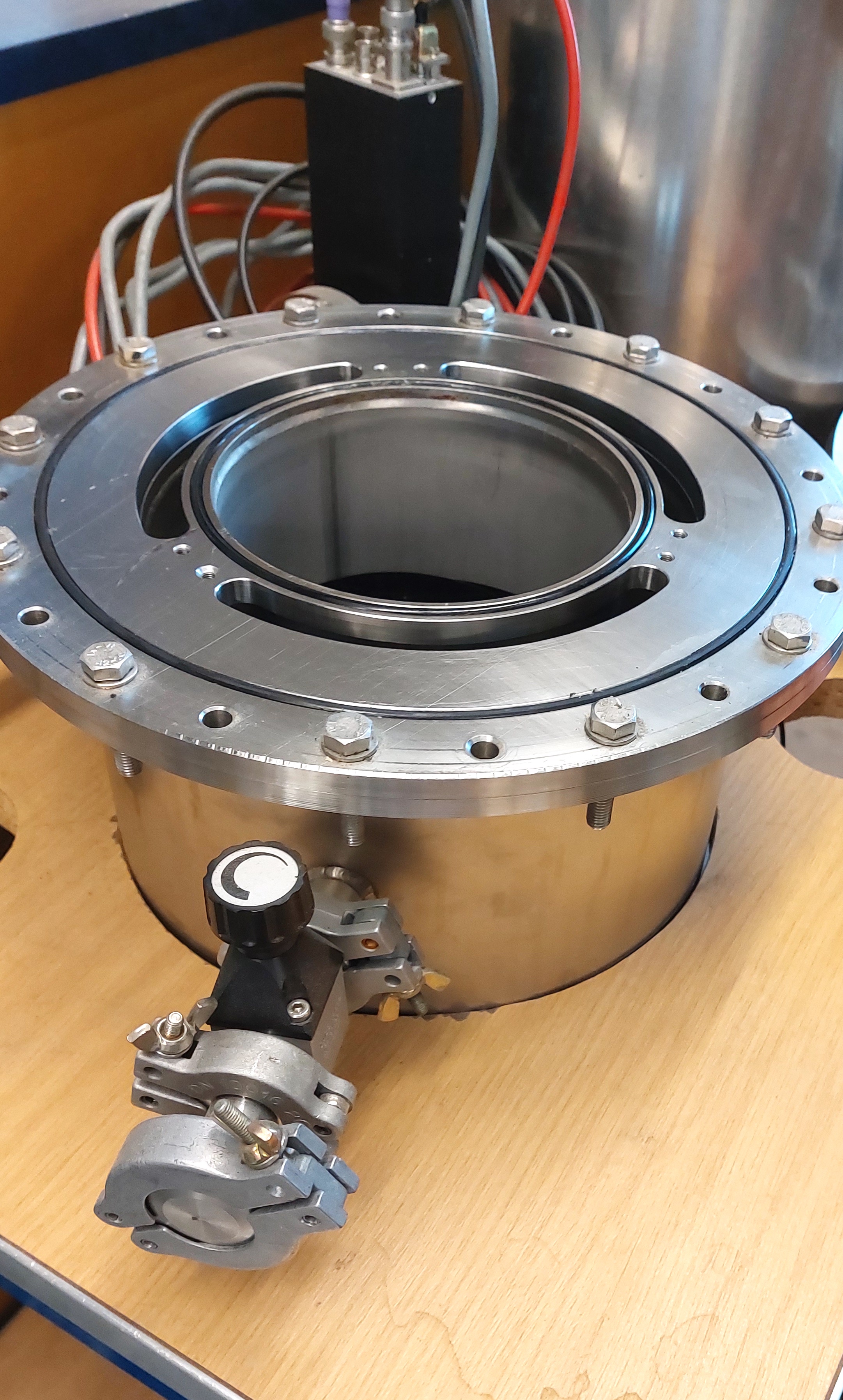}
\caption{(a) Section view of the CAD drawing of the cryostat and the SoLAr prototype. The blue region is indicating the vacuum jacket, the orange region is the middle volume, and the red region is the innermost volume with the prototype detector inside. (b) The cryostat with open lid, where the access to separate volumes is visible.}
\label{fig:cryostat_inside}
\end{figure}

A triple-layer cryostat was used to operate the
SoLAr prototype with liquid argon.
A vacuum jacket, the outermost layer, provides thermal insulation for the two inner volumes of the cryostat. The middle volume is used as a cooling jacket with liquid argon constantly flowing through this layer. The innermost volume of the cryostat is filled with liquid argon once and sealed for the full duration of the run. Figure~\ref{fig:cryostat_inside} shows a schematic of the SoLAr Prototype-v1 in the cryostat. 

The liquid argon passes through a filling filter with copper getters to remove traces of oxygen and humidity form the LAr. This small setup is not equipped with re-circulation capabilities and thus the purity of the LAr will deteriorate over time, giving us a limited time window of about 24 hours to bring up the system and take cosmic-ray data. 

\begin{figure}[htbp]
\centering 
\includegraphics[height= 7.2cm]{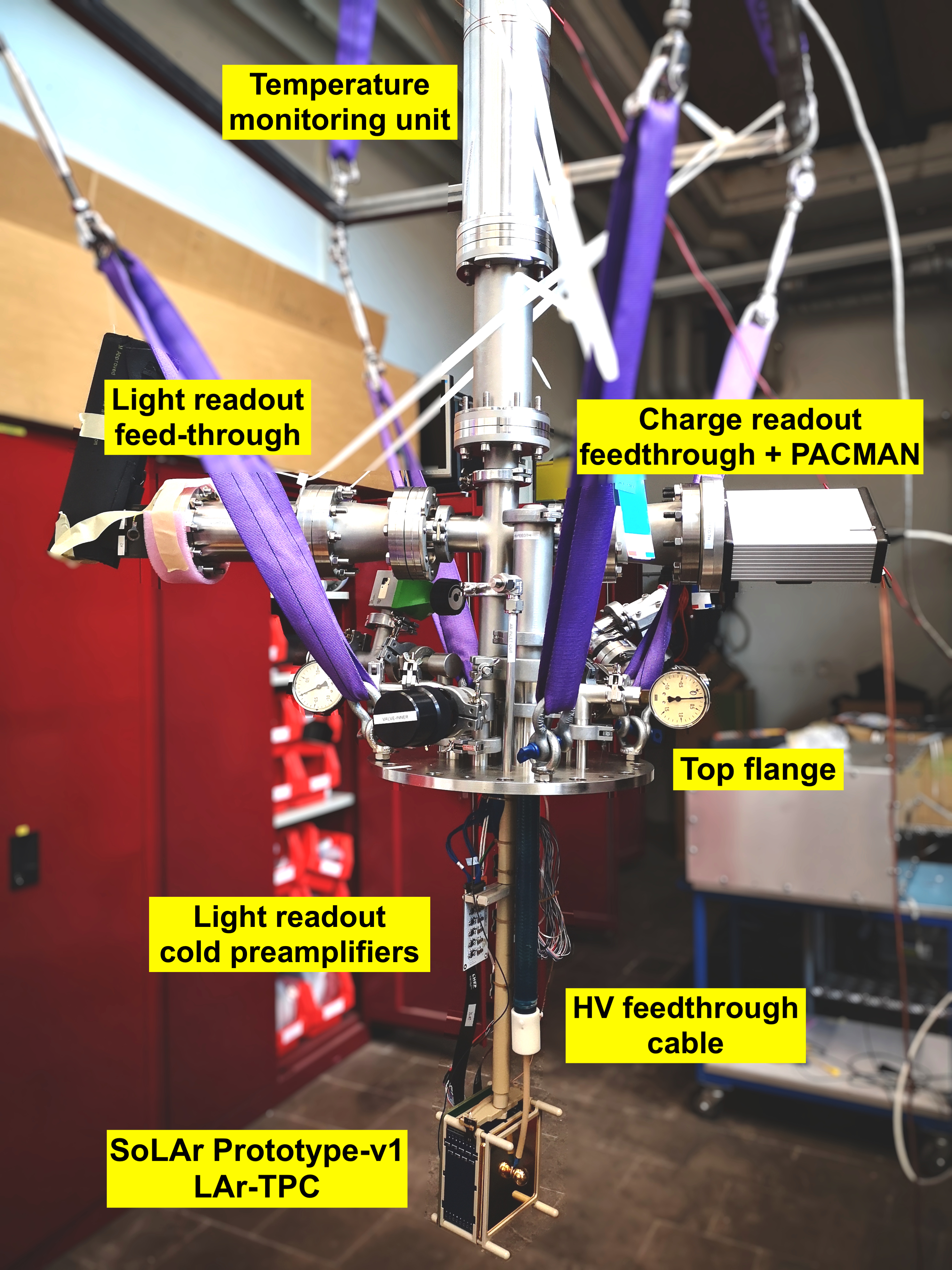}
\includegraphics[height =7.2cm]{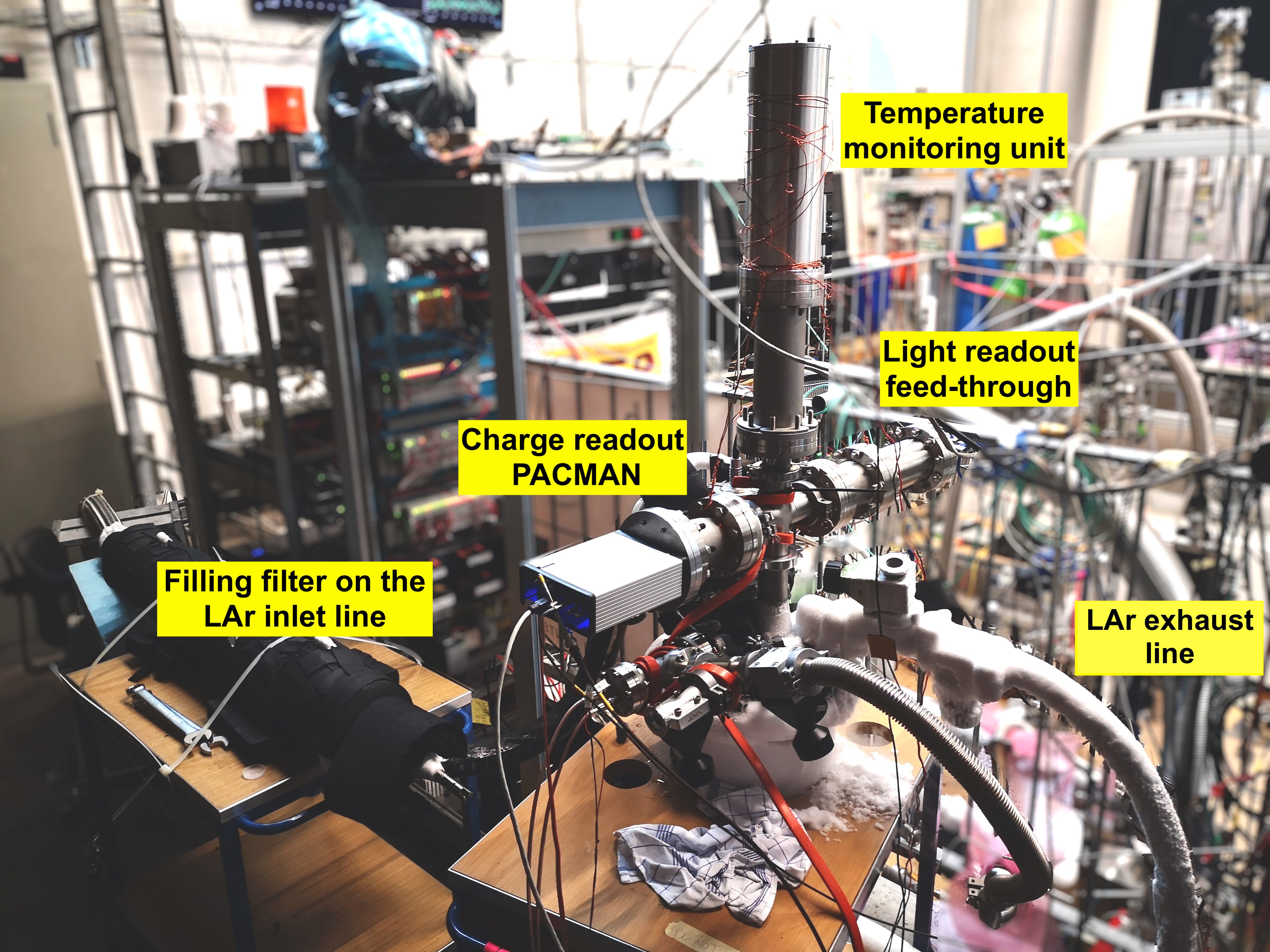}
\caption{\label{fig:top_flange} (a) Top flange with all the feed-throughs and the suspended SoLAr Prototype-v1 LArTPC shortly before inserting it into the cryostat. (b) Experimental setup during cryogenics operation.}
\end{figure}
The level of liquid argon inside the cryostat is controlled with three temperature sensors at different heights connected to the slow control system. The slow control system and each readout system has an individual feed through on the cryostat to transfer data. The bias voltage for the SiPMs is routed through a separate feed through in order to allow different schemes of ground offsets from common ground. 
The pressure in the layers of the cryostat is monitored with two pressure gauges, mounted on the top flange. The inner volume is kept at an overpressure of $50$~mbar for the duration of the test.   
 Figure~\ref{fig:top_flange} shows the top flange with the assembled SoLAr Prototype-v1 TPC shortly before insertion into the cryostat and the experimental setup during cryogenic operation.

%% file: section5_cosmic_run.tex
\section{Results}
\label{sec_run}
\subsection{Prototype Operation}
 In this section, we describe the results obtained with the light and charge readout based on data collected during the cryogenics operation at the University of Bern during October 24--26, 2022. After the cryostat had been filled with liquid argon and the temperature sensors had been submerged, the cathode voltage was raised to a potential of $2.5$~kV to provide an electric field of $500$~V/cm. 

The data was collected for both the charge and the light readout continuously with separate self-triggering data acquisition systems. In order to synchronize, a pulse per second signal from the Global Positioning System (GPS) was fed to both systems. In addition, a light trigger signal was sent to the charge readout PACMAN, which is written into the charge data stream as a $t_0$ marker to identify the start time of an event. This enables correct association of the charge packets' drift time. 14 out of 16 SiPMs were operational and collected full light waveform for each event. Only about 50\% of the charge collecting pixels were working on the low threshold level that is needed. There was no cross talk observed between the working charge collecting pixels and the SiPMs.

The charge and light data have universal time stamps associated with the signals measured on both readout systems. These time stamps are used in offline data processing to associate the light and charge signals, creating what we defined as ``events'', with each event covering a time span of \mbox{$200$~$\mu$s}. 
The drift distance of 
$5$~cm requires $\approx 60$~$\mu$s for an electron to travel the full drift length.
The event window is therefore long enough
so that all the charge that is created through ionization at 
the time $t_0$ can drift from the cathode to the anode.
If two light signals occur in close proximity, only the earliest light signal is used to mark the start of an event. This can potentially cause some event mismatching which has to be taken into account in the analysis.

The scope of this first prototype is the proof of the conceptual idea of having light and charge collection on the anode plane. It should ensure that the new VUV SiPMs are operational in liquid argon on the anode plane and show that it is possible to do light and charge matching for cosmic events.

\subsection{First cosmic-ray tracks with dual-pixel light and charge readout}
\begin{figure}[htbp]
\centering 
\includegraphics[width =0.95\textwidth]{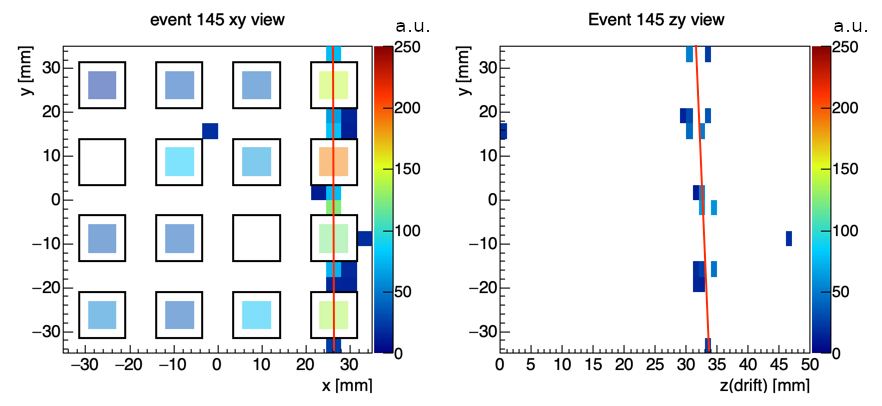}
\includegraphics[width =0.95\textwidth]{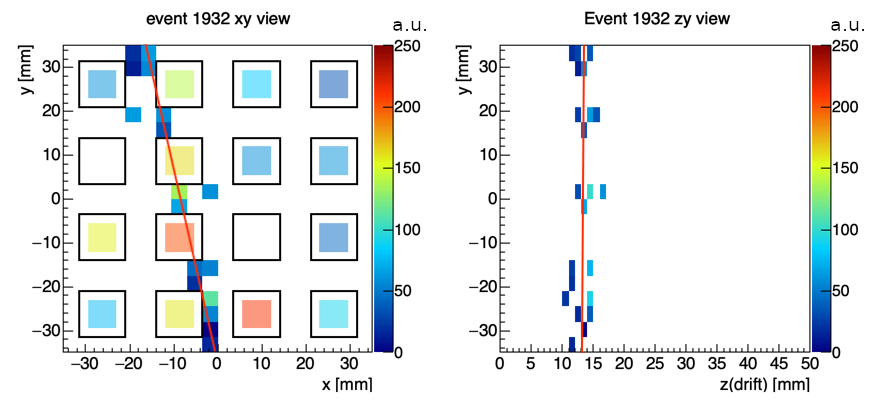}
\quad
\vskip -11.3cm \hskip 5.cm
\includegraphics[height=20mm]{solar_logo_black.png}
\vskip 9.3cm
\vskip -4.7cm \hskip 8.5cm
\includegraphics[height=20mm]{solar_logo_black.png}
\vskip 2.7cm

\caption{Two cosmic-ray events recorded by the prototype shown from the anode view in the $xy$ plane and as $zy$ view, where $z$ is the drift direction.
The $xy$ views show the total light detected by each SiPM in the larger square pads and the amount of collected charge on each pixel in the smaller square pads. 
The color scale represents relative intensity in ADC units. Its range is tuned for visual aid.}
\label{fig:tpc_events}
\end{figure}

An estimated $70,000$ events were collected during the October 2022 run. In Figure \ref{fig:tpc_events}, we show two events with cosmic-ray muon candidates passing through the detector to highlight the association between light and charge information. The $x$ and $y$ axes represent the span of the anode plane, while the $z$ direction indicates the drift distance between the anode and the cathode.
The patterns of charge and light signals on the pixels clearly indicate a muon track candidate. 
The $zy$ projection is obtained by drawing a line of the charge signals to the anode plane, using their light trigger time $t_0$ to calibrate their drift distance to the anode. The position information provides an unambiguous three-dimensional hit object. The track finding occurs by fitting a linear function to the hits in the $xy$ and $xz$ planes with restrictions placed on the minimum number of hits in the event and the $\chi^2$ of the hits relative to the line of best fit in each of the two planes.

This result represents the first combined detection of charge and light on a dual-pixel anode plane in a LArTPC, which is 
a major milestone for demonstrating the fully three-dimensional charge and light association of the SoLAr concept for triggering and reconstruction purposes~\cite{Parsa:2022mnj}.

\subsection{Charge collection per unit length}

\begin{figure}
    \centering
    \includegraphics[width=0.6\linewidth]{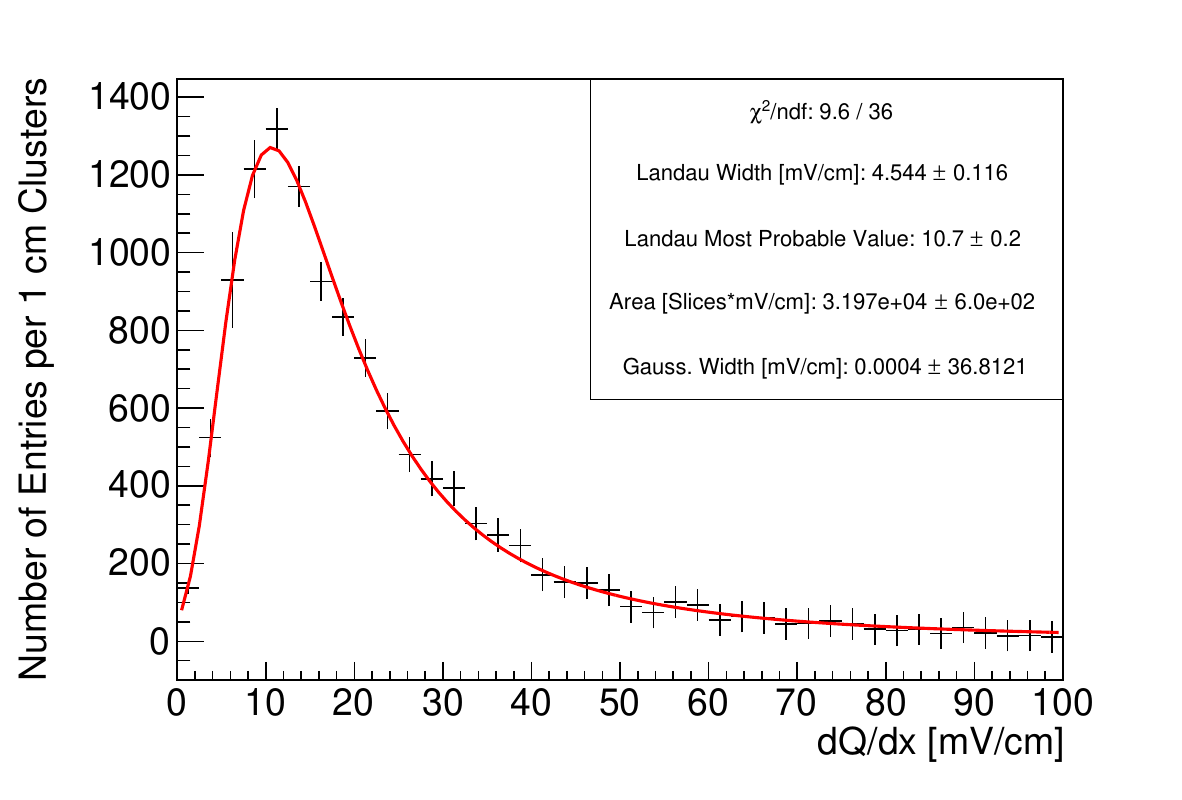}
    \caption{Charge deposit per unit length for through going cosmic-ray particles. Each cluster has a length of $1$~cm. A fit to the histogram using a Landau-Gaussian distribution is also shown~\cite{langauFit}.}
    \label{fig:dqdx}
\end{figure}

We evaluate the amount of charge per unit length ($dQ/dx$) on selected events using through-going tracks selected by applying linear fits to the charge data. These tracks are  analyzed pixel-by-pixel to measure the total amount of charge each pixel records from what is likely a minimally ionizing cosmic-ray muon. 
Figure~\ref{fig:dqdx} shows the distribution of charge per unit length for through-going tracks over a data-taking period of $10$ minutes for the $0$~V ground offset scheme. To ensure that ionization deposits that fall between two pixels are counted correctly, the $dQ/dx$ is measured by combining charge measured by pixels into 1 cm clusters. The value of each individual pixel channel is subtracted by the value of said channel from a previously taken pedestal run to equalize the baseline of said channels. Because approximately half of the pixels were not functional, the purpose of this study is to show that calorimetry with the charge information can be represented by a Landau-Gaussian function. The measurements are presented in ADC/cm. The charge readout system used has been observed to have a 5\% channel-to-channel variation in charge collected, thereby shifting $dQ/dx$~\cite{Madigan:2023fsg}. Therefore, a systematic uncertainty of that scale has been added to the distribution.

As expected, the distribution of electrons per unit length follows a Landau-Gaussian distribution, which is consistent with the result of a fit~\cite{langauFit}. 
The result suggests that the SoLAr Prototoype-v1 charge readout system can carry out basic charge-based calorimetry. 
Further calibration and larger-scale prototypes are necessary to obtain results comparable to similar pixelated prototype detectors~\cite{module0} since the volume of the field shell and 
the active detector area for charge collection are small, leading 
to charge loss.

\subsection{Light collection as a function of SiPM ground offset voltages }

The setup is designed with the possibility to put the ground voltage level of the SiPMs to an offset with respect to the ground level on the anode surface and the charge collection pads. In this scheme, the voltage level of both pins of the SiPMs is simultaneously shifted by a fixed amount while the SiPM bias, the difference between potential of the two SiPM pins, is kept at the same value of $46$~V. 

\begin{figure}
    \centering
\includegraphics[width=0.7\textwidth]{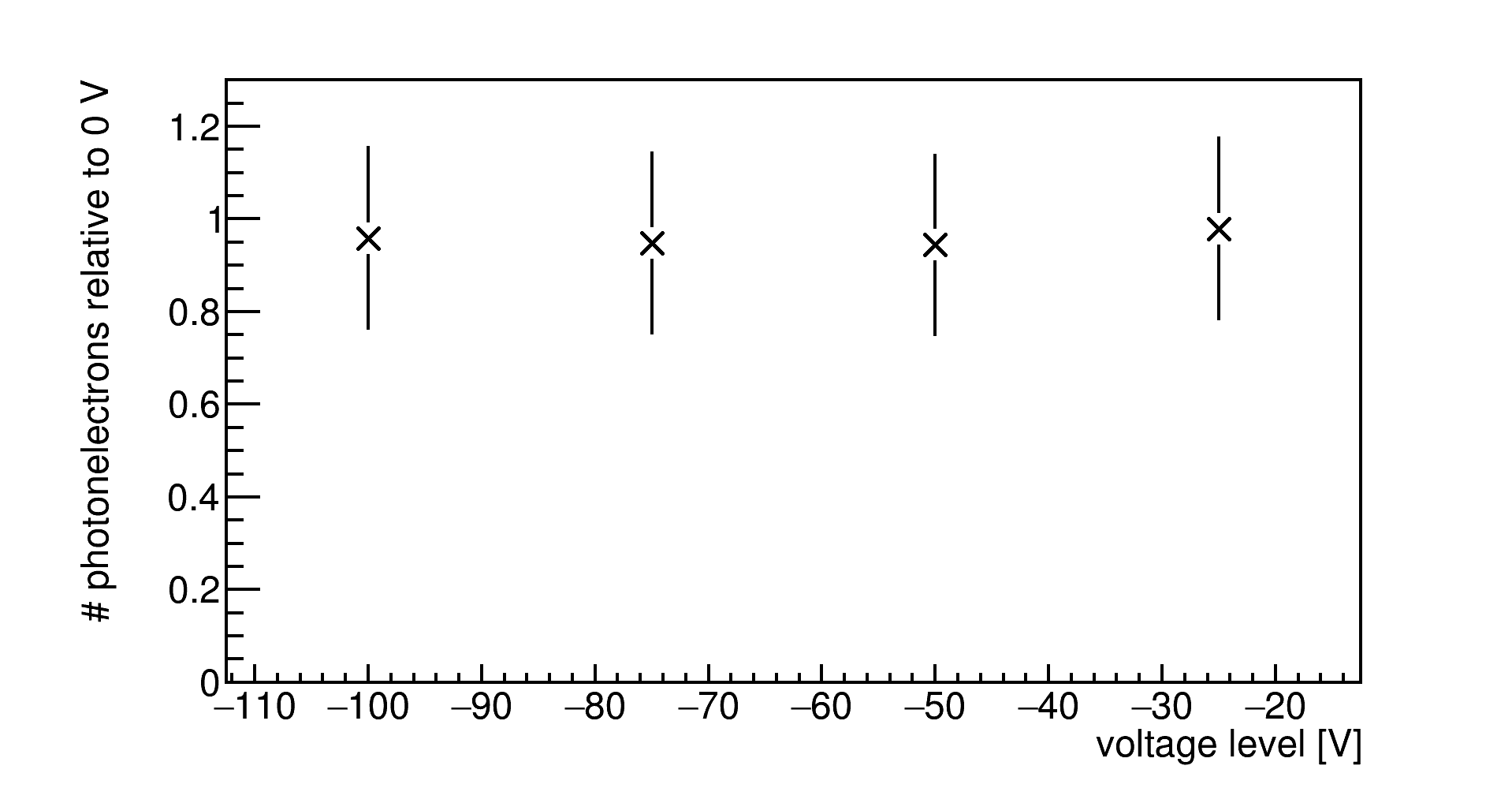}
    \caption{Average light yield detected per unit length of muon tracks as a function of the SiPM ground offset voltage, relative to the average light yield with no ground offset voltage applied. The uncertainties shown are statistical only.}
    \label{fig:floating}
\end{figure}

Putting the SiPMs to a negative potential ground offset may increase the effective charge collection of neighboring pads by deforming the electric field lines and thus diverting the drifting electrons away from the SiPMs surface and to the charge collection pixels (see Figure~\ref{fig:comsol}). The potential ground offset should not have an impact on the performance of the SiPMs because the potential difference within the SiPM is kept constant. We study the amount of collected light as a function of the ground offset voltage applied to the SiPMs by comparing the average number of detected photons per unit length of through-going muon tracks for different ground offset voltage levels.

Figure~\ref{fig:floating} shows the number of photoelectrons collected for each ground offset level relative to the mean number of collected photoelectrons without any ground offset voltage applied to the SiPMs. The calibrated gain is not expected to change and is therefore not reevaluated, as the electronics remain the same across different voltages, and in particular the voltage difference within the SiPM is kept constant. For each voltage ground offset, we collected a total of about 5,000 events, except for the setting $-25$~V where it is approximately 3,000 events. As expected, the mean signal strength as a function of ground offset voltages is constant within the statistical uncertainties shown.

The data collected with the charge pads did not meet our requirement for obtaining information on the charge collection performance at various ground offset levels. Additionally the SiPMs are surrounded by a ceramic case which increases the distance between the SiPM and the charge-collecting pixels. This makes it difficult to obtain representative results for future designs that will not have ceramic casings around the SiPMs.

%% file: section6_conclusion.tex
\section{Summary}
We operated for the first time a liquid-argon TPC with a new readout concept (SoLAr), integrating charge and light sensors on the same surface, to perform native 3D reconstruction of an event. In this paper we present the design of a first small demonstrator, referred to as SoLAr Prototype-v1, which used an integrated dual-pixel readout of light and charge pixels, based on a distributed arrangement of VUV SiPMs on the anode plane. Prototype-v1 was operated in October 2022 at the University of Bern and collected cosmic-ray muon tracks. We show unambiguous correlations between the light and charge information from these cosmic-ray tracks. We also show that the charge collection of the ionization electrons corresponds to a Landau-Gaussian distribution, demonstrating that this detector has the potential to perform calorimetric measurements despite the interleaved SiPMs. The light collection on the SiPM is efficient and independent on the ground offset voltage of the SiPM bias, necessary to tune the drift field to guide the ionization electrons to the charge pads.
The results from this first prototype opens the path for the design of detectors using the integrated light-charge pixel concept to detect interactions from GeV to the MeV energy scale in very large (multi-kiloton) mass detectors, suitable e.g. for solar neutrino measurements. We are preparing next prototyping stages to scale the size and characterize and optimize the capability of the SoLAr concept for LArTPCs to detect neutrino signals at the MeV-scale. 

%% file: main.bbl
\providecommand{\href}[2]{#2}\begingroup\raggedright\begin{thebibliography}{10}

\bibitem{DUNETDRVol1}
{\scshape DUNE} collaboration, B.~Abi et~al., \emph{{Deep Underground Neutrino
  Experiment (DUNE), Far Detector Technical Design Report, Volume I
  Introduction to DUNE}},
  \href{http://dx.doi.org/10.1088/1748-0221/15/08/T08008}{\emph{JINST}
  {\bfseries 15} (2020) T08008},
  [\href{https://arxiv.org/abs/2002.02967}{{\ttfamily 2002.02967}}].

\bibitem{DUNE:2020jqi}
{\scshape DUNE} collaboration, B.~Abi et~al., \emph{{Long-baseline neutrino
  oscillation physics potential of the DUNE experiment}},
  \href{http://dx.doi.org/10.1140/epjc/s10052-020-08456-z}{\emph{Eur. Phys. J.
  C} {\bfseries 80} (2020) 978},
  [\href{https://arxiv.org/abs/2006.16043}{{\ttfamily 2006.16043}}].

\bibitem{2004HEPReview}
K.~Kubodera and T.-S. Park, \emph{{The Solar HEP process}},
  \href{http://dx.doi.org/10.1146/annurev.nucl.54.070103.181239}{\emph{Ann.
  Rev. Nucl. Part. Sci.} {\bfseries 54} (2004) 19--37},
  [\href{https://arxiv.org/abs/nucl-th/0402008}{{\ttfamily nucl-th/0402008}}].

\bibitem{SNO2020}
{\scshape SNO} collaboration, B.~Aharmim et~al., \emph{{Search for $hep$ solar
  neutrinos and the diffuse supernova neutrino background using all three
  phases of the Sudbury Neutrino Observatory}},
  \href{http://dx.doi.org/10.1103/PhysRevD.102.062006}{\emph{Phys. Rev. D}
  {\bfseries 102} (2020) 062006},
  [\href{https://arxiv.org/abs/2007.08018}{{\ttfamily 2007.08018}}].

\bibitem{Capozzi}
F.~Capozzi, S.~W. Li, G.~Zhu and J.~F. Beacom, \emph{{DUNE as the
  Next-Generation Solar Neutrino Experiment}},
  \href{http://dx.doi.org/10.1103/PhysRevLett.123.131803}{\emph{Phys. Rev.
  Lett.} {\bfseries 123} (2019) 131803},
  [\href{https://arxiv.org/abs/1808.08232}{{\ttfamily 1808.08232}}].

\bibitem{IGil-Botella_2004}
I.~Gil~Botella and A.~Rubbia, \emph{{Decoupling supernova and neutrino
  oscillation physics with LAr TPC detectors}},
  \href{http://dx.doi.org/10.1088/1475-7516/2004/08/001}{\emph{JCAP} {\bfseries
  08} (2004) 001}, [\href{https://arxiv.org/abs/hep-ph/0404151}{{\ttfamily
  hep-ph/0404151}}].

\bibitem{Fogli:2006fu}
G.~L. Fogli, E.~Lisi, A.~Marrone and A.~Palazzo, \emph{{Solar neutrinos: With a
  tribute to John. N. Bahcall}},  in \emph{{3rd International Workshop on
  NO-VE: Neutrino Oscillations in Venice: 50 Years after the Neutrino
  Experimental Discovery}}, pp.~69--80, 5, 2006.
\newblock \href{https://arxiv.org/abs/hep-ph/0605186}{{\ttfamily
  hep-ph/0605186}}.

\bibitem{DUNE:2020zfm}
{\scshape DUNE} collaboration, B.~Abi et~al., \emph{{Supernova neutrino burst
  detection with the Deep Underground Neutrino Experiment}},
  \href{http://dx.doi.org/10.1140/epjc/s10052-021-09166-w}{\emph{Eur. Phys. J.
  C} {\bfseries 81} (2021) 423},
  [\href{https://arxiv.org/abs/2008.06647}{{\ttfamily 2008.06647}}].

\bibitem{Parsa:2022mnj}
S.~Parsa et~al., \emph{{SoLAr: Solar Neutrinos in Liquid Argon}},
  {\emph{{Snowmass 2021}} (2022) },
  [\href{https://arxiv.org/abs/2203.07501}{{\ttfamily 2203.07501}}].

\bibitem{DUNETDRVol4}
{\scshape DUNE} collaboration, B.~Abi et~al., \emph{{Deep Underground Neutrino
  Experiment (DUNE), Far Detector Technical Design Report, Volume IV: Far
  Detector Single-phase Technology}},
  \href{http://dx.doi.org/10.1088/1748-0221/15/08/T08010}{\emph{JINST}
  {\bfseries 15} (2020) T08010},
  [\href{https://arxiv.org/abs/2002.03010}{{\ttfamily 2002.03010}}].

\bibitem{DUNE:2023nqi}
{\scshape DUNE} collaboration, A.~Abed~Abud et~al., \emph{{The DUNE Far
  Detector Vertical Drift Technology. Technical Design Report}},
  \href{http://dx.doi.org/10.1088/1748-0221/19/08/T08004}{\emph{JINST}
  {\bfseries 19} (2024) T08004},
  [\href{https://arxiv.org/abs/2312.03130}{{\ttfamily 2312.03130}}].

\bibitem{duneNDCDR}
{\scshape DUNE} collaboration, V.~Hewes et~al., \emph{{Deep Underground
  Neutrino Experiment (DUNE) Near Detector Conceptual Design Report}},
  \href{http://dx.doi.org/10.3390/instruments5040031}{\emph{Instruments}
  {\bfseries 5} (2021) 31}, [\href{https://arxiv.org/abs/2103.13910}{{\ttfamily
  2103.13910}}].

\bibitem{DUNE:2022aul}
{\scshape DUNE} collaboration, A.~Abed~Abud et~al., \emph{{Snowmass Neutrino
  Frontier: DUNE Physics Summary}},
  \href{https://arxiv.org/abs/2203.06100}{{\ttfamily 2203.06100}}.

\bibitem{dune2}
{\scshape DUNE} collaboration, A.~A. Abud et~al., \emph{{DUNE Phase II:
  Scientific Opportunities, Detector Concepts, Technological Solutions}},
  \href{https://arxiv.org/abs/2408.12725}{{\ttfamily 2408.12725}}.

\bibitem{Dwyer:2018phu}
D.~A. Dwyer et~al., \emph{{LArPix: Demonstration of low-power 3D pixelated
  charge readout for liquid argon time projection chambers}},
  \href{http://dx.doi.org/10.1088/1748-0221/13/10/P10007}{\emph{JINST}
  {\bfseries 13} (2018) P10007},
  [\href{https://arxiv.org/abs/1808.02969}{{\ttfamily 1808.02969}}].

\bibitem{comsol}
{COMSOL Multiphysics {\textcopyright} v.6.2, COMSOL AB, Stockholm, Sweden}.
  {{\url{https://www.comsol.com}}}.

\bibitem{Madigan:2023fsg}
P.~Madigan, \emph{{Measurement of Muon Capture on Argon with a Pixelated Liquid
  Argon Time Projection Chamber}}.
\newblock PhD thesis, University of California, Berkeley, 2023.
\newblock https://escholarship.org/uc/item/7k13v4b0.

\bibitem{langauFit}
H.~Pernegger and M.~Friedl, ``{ROOT reference guide: Convoluted Landau and
  Gaussian Fitting Function}.''
  {\url{https://root.cern/doc/master/langaus_8C.html}}.

\bibitem{module0}
{\scshape DUNE} collaboration, A.~Abed~Abud et~al., \emph{{Performance of a
  modular ton-scale pixel-readout liquid argon time projection chamber}},
  \href{https://arxiv.org/abs/2403.03212}{{\ttfamily 2403.03212}}.

\end{thebibliography}\endgroup
